\documentclass[12pt]{iopart}

\begin{document}

\title{Efficient search by optimized intermittent random walks}

\author{Gleb Oshanin$^{1,2}$, Katja Lindenberg$^3$, Horacio S Wio$^4$ and Sergei Burlatsky$^5$ }
\address{$^1$Laboratory J.-V. Poncelet (UMI  CNRS 2615),
Independent University of Moscow, Bolshoy Vlasyevskiy Pereulok 11, 119002 Moscow, Russia}
\address{$^2$Laboratoire de Physique Th\`eorique de la Mati\`ere Condens\'ee (UMR 7600), Universit\'e Pierre et Marie Curie/CNRS, 4
place Jussieu, 75252 Paris Cedex 5 France}
\address{$^3$Department of Chemistry and
Biochemistry, University of California, San Diego, La Jolla, CA
92093-0340 USA}
\address{$^4$Instituto de F\'{\i}sica de
Cantabria, Avda. Los Castros s/n, E-39005 Santander, Spain}
\address{$^5$United Technologies
Research Center, United Technologies Corp., 411 Silver Lane, 129-21
East Hartford, CT 06108, USA}
\ead{oshanin@lptmc.jussieu.fr; klindenberg@ucsd.edu; wio@ifca.unican.es; burlatsf@utrc.utc.com}

\begin{abstract}
We study the kinetics for the search of an immobile target by randomly moving searchers that
detect it only upon encounter.
The searchers perform intermittent random walks on a one-dimensional lattice.
Each searcher can step on a nearest neighbor site
with probability $\alpha$, or go off lattice
with probability $1 - \alpha$ to
move in a random direction until it lands back on the lattice at a
fixed distance $L$ away from the departure point.
Considering $\alpha$ and $L$ as optimization parameters,
we seek to enhance the chances of successful detection by
minimizing the probability $P_N$ that the target remains undetected up to
the maximal search time $N$.
We show that even in this simple model a number of very
efficient search strategies can lead to a decrease of
$P_N$ by orders of magnitude upon appropriate choices of $\alpha$ and $L$.
We demonstrate that, in general, such optimal
intermittent strategies are much more efficient than Brownian searches and
are as efficient as search algorithms based on random walks with heavy-tailed
Cauchy jump-length distributions.  In addition, such intermittent strategies
appear to be more advantageous than L\'evy-based ones
in that they lead to more thorough exploration of visited regions in space and
thus lend themselves to parallelization of the search processes.
\end{abstract}

\pacs{05.40.-a, 87.23.-n}
\vspace{2pc}
\noindent{\it Keywords}: search processes, optimization, intermittent random walks

\vspace{2pc}
\submitto{\JPA}
\maketitle

\section{Introduction}

Search processes are ubiquitous in nature:
Predators search for prey, prey also hunt,
molecules search for each other to recombine
in order to produce required chemicals, proteins
search for target sequences on DNAs.
Human beings spend their lives searching for different things -
better jobs, shelters, partners, lost keys;
they also seek efficient search strategies
to minimize the time to reach the desired target,
or at least to enhance their chances of eventually finding it.

The search for a desired target may depend on a variety of different
conditions and may take place in different environments. Targets may be
sparse, hidden, difficult to detect even when found. The targets may
be mobile or immobile, they may try to avoid searchers, there
may be one target or many. They may have a finite
life-time and vanish before they are detected. Searchers, on the other
hand, may search ``blindly," detecting the target only upon encounter,
or they may perceive distant targets and adjust their
motion accordingly. They may have no memory of previously visited areas, or
they may avoid such areas.  The searchers may act
individually or in swarms, optimizing their search efficiency by exchanging
information. Finally, the ``efficiency" of a search may be judged by a variety of measures,
including the time to reach a target or targets, the number of encounters of searchers and targets
per unit time, or the exploration range of space per unit time.
In general, for each specific situation different search
strategies may be appropriate.  The quest for optimal strategies
has motivated a great deal of
work in the past few
years~\cite{stephens,stone,bell,frost,klafter,viswanathan,viswanathan2,levandovsky,bartumeus,boyer,bartumeus2,boyer2,obrien,kramer,benichou1,benichou2,benichou3,coppey,reynolds,1,yossi,sub1,sub2,sub3,sub4,sub5,james}.

Earlier work tended to focus on deterministic
search algorithms (see, e.g., Refs.\cite{stephens,stone,bell,frost} and references therein)
specific to human activities such as, say, rescue operations
or the search for natural resources.
More recent studies have focused on random search strategies.
In this context, it has become quite clear
that strategies based
on L\'evy flights (instantaneous) or L\'evy walks (occurring with a finite velocity)
are according to all measures more advantageous
than strategies based on conventional diffusive motion or random
walks with steps to nearest neighbors only (Brownian
search)~\cite{klafter,viswanathan,viswanathan2,levandovsky,bartumeus,boyer,bartumeus2,boyer2}.
L\'evy searchers perform excursions whose lengths $l$  are random variables with heavy-tailed distributions
$p(l)$
of
the form
\begin{equation}
\label{levy}
p(l) \sim \frac{B}{l^{\mu + 1}}, \;\;\; 0 < \mu < 2,
\end{equation}
where $B$ is a normalization constant.
In particular, it was demonstrated in Ref.~\cite{szu} that using the
Cauchy distribution of jump lengths as given in equation~(\ref{levy}) with $\mu = 1$
instead of a Gaussian distribution (Brownian search) allows for a faster
cooling scheme, and hence for a considerable reduction of computer
time in the search for a global minimum in nonconvex (multiple extrema)
energy landscapes by simulated annealing. Aside from this,
extensive data has been presented allegedly supporting the idea that many of
species in the living world do indeed follow L\'evy-type random
motion in their
search~\cite{klafter,viswanathan,viswanathan2,levandovsky,bartumeus,boyer,bartumeus2,boyer2}.
This point, however, has been questioned recently in Refs.~\cite{edwards, travis}. The main
objections of Refs.~\cite{edwards, travis} have recently been re-examined in Ref.~\cite{boyer2},
where it was concluded that L\'evy-based strategies may still be consistent with
experimental observations if one takes into account a highly non-homogeneous spatial
distribution of prey. More sophisticated models with adaptive
behavior in which the foragers use their cognitive skills to develop more efficient
foraging strategies have also started to appear in the literature (see, e.g., Ref.~\cite{boyer3}).

Following the observation of trajectories of foraging animals such
as lizards or fish or birds, in which
active local search phases randomly alternate with relocation phases (see, e.g.,
Refs.\cite{obrien,kramer}),
another type of random search - an intermittent search - has been proposed. In this algorithm, a search
is characterized by two distinct
types of motion, a ballistic
relocation stage during which the searcher is non-receptive to the target, and a relatively slow phase
with random Brownian-type motion during which the target may be
detected~\cite{benichou1,benichou2,benichou3,coppey,1} (see also Ref.~\cite{reynolds}). In this approach
one aims to minimize the time of first passage to the target from a given location by varying, for
example, the relative durations of the active and relocation stages.

In this paper we pursue the optimization of an intermittent strategy.
A simpler version of our model was presented in Ref.~\cite{1}, where
we developed a search algorithm based on intermittent random walks that involve nearest neighbor
steps and off-lattice relocations of a fixed length $L$.
There, instead of minimizing the first passage time to the target, we sought to
maximize the success of the search by minimizing the target non-detection probability
\emph{over a fixed finite maximum search time}. We note that the time derivative
of this probability defines the distribution function of the first passage time to
the detection event. Thus, contrary to previous work
in which only the mean first passage time was
optimized~\cite{benichou1,benichou2,benichou3,coppey,1},
our goal was to optimize the full distribution function.
Our optimization parameter was the intermittency parameter $\alpha$ that determines whether the next
step will be on- or off-lattice.  It was shown, both analytically and numerically,
that the probability of failure to detect the target over a finite search time
can be made smaller by many orders of magnitude upon an appropriate choice of
this parameter. We note that in Ref.~\cite{yossi} a different intermittent search
algorithm was proposed in which the length of the relocation stage is a random variable
with a heavy-tailed distribution in equation~(\ref{levy}). In that work it was concluded that such
a combined strategy is advantageous over intermittent searches with
exponentially distributed~\cite{benichou1,benichou2,benichou3} or fixed~\cite{1} relocation lengths
since it allows a searcher to find the target more quickly in the critical case of rare targets, and since
the search performance is much less dependent on adaptation to the target density.
However, we argue that in some (albeit not all) situations, strategies involving L\'evy
distributed relocations can not be optimal. This occurs when there is some maximal time that
the search is allowed to run. Allowing the length of the relocation stage (and
consequently, the time spent on each relocation tour) to have a
heavy-tailed distribution would lead to some portion of the search process that would
involve unnecessarily long relocations divorced from the targets, thus not contributing to the
overall finite time effort. Thus, when appropriate, an
optimal search algorithm should be based on relocation length distributions that explicitly
account for the fact that a search process is limited in time.
It may well be that the optimal jump-length distribution
should itself vary with time.
Furthermore, trajectories of L\'evy walks or flights are ``overstretched" in the sense that such walks
explore space in a very irregular manner. The visited area
consists of a patchy set of disconnected clusters, leaving large
unexplored voids compared to the case of a Brownian search.
Additionally, when many L\'evy searchers are involved, the fact that a
L\'evy distribution does not have moments induces rapid
mixing. This mixing might be advantageous if the detection
probability is low (or the false alarm probability high), such
that multiple rechecking of each site by other searchers is
required in order to spot the target. Otherwise, this very efficient
mixing might be a disadvantage since it
does not favor the parallelization of the search process by
dividing the searched area into subunits and assigning a separate domain to each
searcher. In the ``living" world animals are often
constrained to their assigned territories, and even an occasional
incursion into a neighbor's terrain while searching for prey may
cause serious difficulties.

In this paper we revisit the question of an optimal
jump-length distribution underlying an efficient search algorithm.
Focussing on the one-dimensional case (for which L\'evy-based search strategies are
said to most dramatically outperform intermittent ones),
we study the search kinetics of a ``hidden" immobile target located at the origin of
an infinite lattice by a concentration $\rho$ of randomly moving searchers.
The motion of the searchers is intermittent, consisting of two distinct, randomly alternating
stages - ballistic, off-lattice
relocations with finite velocity over a fixed distance $L$,
and random walks between nearest-neighboring sites.
In other words, we consider a search by
intermittent random walkers  with a jump-length distribution
of the form [very different from that in equation~(\ref{levy})]
\begin{eqnarray}
\label{Int}
p(l) = \frac{\alpha}{2} \left[\delta(l - 1) + \delta(l + 1)\right]
+ \frac{(1 - \alpha)}{2} \left[\delta(l - L) + \delta(l + L)\right],
\end{eqnarray}
such that the searchers step on nearest neighbors with probability $\alpha$ and,
with probability $1 - \alpha$, perform long jumps over a distance $L$~\cite{1}.
The process evolves in discrete time $n = 0,1,2, \ldots,N$, where $N$ is
the maximal time the search process may run.
This time may depend, for instance, on our patience or on experimental constraints.
Note that the constraint of the maximal search time $N$ is the crucial aspect of our work
which makes our analysis very different from other models.
Steps to nearest neighbors take unit time, while off-lattice relocations over a
distance $L$ require time $T$.  The term ``hidden" means that a searcher
can not perceive the target when off-lattice.  A searcher only detects the target
when it arrives at the site on which the target is located.

We pose the following question: Is it possible to choose
$\alpha = \alpha_N$ and $L=L_N$, dependent on the maximal search
time $N$ but independent of the running time $n$, which
optimizes the search efficiency and leads to a performance that is better than
L\'evy-based strategies? In order to answer our question, our first goal is to
calculate the probability $P_N$ that the target remains undetected
up to the maximal search time $N$ and to determine its asymptotic
behavior analytically in the large-$N$ limit. Then, considering
$\alpha = \alpha_N$ and $L = L_N$ as optimization parameters, we
seek to enhance the searchers' chances of success by
minimizing $P_N$. We will demonstrate that, depending on
whether we are at liberty to tune $\alpha$ (as in Ref.~\cite{1}), or both $\alpha$ and $L$,
different optimal strategies can be realized all of which
can decrease the value of $P_N$ by many orders of
magnitude compared to a Brownian search. We also
show that even the simple distribution~(\ref{Int})
with optimal $\alpha = \alpha_N$ and $L = L_N$ yields very
efficient search algorithms comparable to and in some cases better than
L\'evy-based
strategies.  Moreover, in striking contrast to the latter, optimal intermittent walks
lead to much denser exploration of space.
These results support our claim that L\'evy-based
searches are not the best algorithms when the search is limited in time.

\section{Model and basic equations}
Consider a one-dimensional regular lattice of unit spacing
containing $M$ sites labeled by $s$. The lattice is a circle, that is, we use periodic boundary conditions.
At one of the lattice sites, say at the
origin $s = 0$, we place an immobile target.  Then we randomly place $K$ searchers
under the constraint that none is placed at the site of the hidden target.
We focus on the behavior in the thermodynamic limit, $K \to \infty$, $M \to  \infty$, with
a finite mean density of searchers $\rho = K/M$. We note that
the model under study can also be solved in the general case of finite $K$ and $M$, but the calculations
become more involved without adding significant new features to our conclusions.

Next we allow the searchers to
move according to the following rule.
At each tick of the clock, $n =1,2,3, \ldots, N$,
each searcher selects randomly between two possibilities:
with probability $\alpha$, it moves with equal likelihood to one of its nearest
neighboring sites, and with probability $(1 - \alpha)$ it leaves the
lattice and flies off-lattice with a given velocity $V$
until it lands $L$ sites away from the departure site. The distance $L$ is fixed,
but either direction of the flight is chosen at
random with equal probabilities. The time it spends
off-lattice during the flight is $T = L/V$. This parameter
can take integer values, $T = 1,2, \ldots, L$. Note that this condition
defines the velocity $V$. There is no restriction on multiple searcher occupancy of the sites.
Note as well that in the two ``pure" cases, $\alpha = 1$ and $\alpha =
0$, the model reduces to standard random walks.
Here we take both $\alpha$ and $L$ to be independent of $n$
but possibly dependent on the maximal search time $N$. An interesting
situation with time-dependent $\alpha = \alpha_n$ and $L = L_n$ will
be discussed elsewhere~\cite{2}.
We focus on perfect detection, that is, a searcher recognizes the target immediately upon
first contact.  The case of imperfect recognition can be solved using the same techniques and will
also be discussed elsewhere~\cite{2}.

The probability that the target has not been detected by step $N$ is related in a simple way to
$S_N$, the \emph{number of distinct sites} visited up to that time,
as
\begin{equation}
\label{survival}
P_N = \exp\left(- \rho S_N\right)
\end{equation}
(see, e.g., Ref.~\cite{hughes}).
This result, which holds for independent searchers and was explicitly shown to be valid for our model
in~\cite{1},
is a crucial equation since we will arrive at results for $P_N$
via calculations of $S_N$.
A larger $S_N$ leads to a  smaller probability that the target
remains undetected and thus to a better search algorithm.
In general, $S_N \sim A N^{\gamma}$ with $0 < \gamma
\leq 1$.  One thus expects that larger $\gamma$ leads to a more efficient
search and explains why, intuitively, it was believed that
the most efficient search algorithm should be based on L\'evy walks
with a broad distribution of jump lengths, for which
$S_N$ grows more rapidly than in the case of simple Brownian motion.
As an aside, however, we note that
even for standard Brownian motion
$S_N \sim A N/\ln(N)$ in two dimensions and $S_N \sim A N$ in three dimensions; consequently,
choosing L\'evy walks as
a search mechanism will not lead to any significant gain
compared to a Brownian search in these higher dimensions
(except perhaps through the prefactor $A$).
In one dimension, however, there are significant differences in
the growth rates of $S_N$ between L\'evy and Brownian motions,
and our task is to explore the place of intermittent random walks in this panorama.

\section{Expected number of distinct visited sites}

The expected number of distinct sites visited can be calculated once we determine
the probability $P(s|s_0;n)$ that a given searcher, starting its intermittent random walk at
site $s_0$ at time moment $n = 0$, appears at site $s$ at time moment $n$. More specifically,
the generating  function $S(z)$ of $S_N$, defined as
\begin{equation}
\label{d}
S(z) = \sum_{N=0}^{\infty} S_N z^N,
\end{equation}
and the lattice Green function (or site occupation generating function)
\begin{equation}
P(s|s_0;z) = \sum_{n=0}^{\infty} P(s|s_0;n) z^n
\end{equation}
of the intermittent random walk are related to each other through~\cite{1}
\begin{equation}
\label{def}
S(z) = \frac{1}{1 - z} \frac{\sum_{s} P(s|0;z)}{P(0|0;z)}.
\end{equation}
Hence, given $P(s|0;z)$, we obtain $S(z)$ by virtue of equation~(\ref{def}).
We then determine the $N$-dependence of  $S_N$ by inverting the discrete Laplace
transform in equation~(\ref{d}).

The probability $P(s|s_0;n)$ obeys the recurrence relation
\begin{eqnarray}
\label{r}
P(s|s_0;n) &=& \frac{\alpha}{2} \left[P(s-1|s_0;n-1) + P(s +
1|s_0;n-1)\right] + \nonumber\\
&+&  \frac{1- \alpha}{2} \left[P(s-L|s_0;n - T) + P(s + L|s_0;n -
T)\right],
\end{eqnarray}
which explicitly takes into account that jumps between
nearest-neighboring sites proceed in unit time, while long-range
jumps over distance $L$ require an integer time $T$.
Equation~(\ref{r}) thus defines a non-Markovian process with a memory.
Note also that since the intermittent random walk
defined by equation~(\ref{r}) is homogeneous so that
$P(s|s_0;n) = P(s - s_0|0;n)$, without loss of generality we henceforth set $s_0 = 0$.

Multiplying both sides of  equation~(\ref{r}) by $z^n$ and performing the summation, we
find that $P(s|0;z)$
obeys
\begin{eqnarray}
\label{2}
P(s|0;z) = \frac{1}{\pi} \int^{\pi}_{0} \frac{\cos\Big(k s\Big) dk}{1 - \alpha z \cos(k) - (1 - \alpha) z^T \cos(k L)}
\end{eqnarray}
and consequently the generating function of the expected number of
distinct visited sites is given by
\begin{eqnarray}
\label{s}
\fl S(z) = \frac{\pi}{(1 - z) \left(1 - \alpha z - (1 - \alpha) z^T \right)}
 \left[ \int^{\pi}_{0} \frac{dk}{1 - \alpha z \cos(k) - (1 - \alpha) z^T \cos(k L)}\right]^{-1}.
\end{eqnarray}
Before we proceed further, the following remarks are in order.
Note that $P(s|0;z)$ in equation~(\ref{2}), and consequently
$S(z)$ in equation~(\ref{s}), can be calculated explicitly in the two ``pure" random walk cases,
$\alpha = 1$ and $\alpha = 0$. In these two limits one finds for $S_N$ the
large-$N$ asymptotic behavior
\begin{equation}
\label{rw1}
S_N(\alpha = 1) = \left(\frac{8 N}{\pi}\right)^{1/2} + O\left(\frac{1}{N^{1/2}}\right)
\end{equation}
and
\begin{equation}
\label{rw2}
S_N(\alpha = 0) = \left(\frac{8 N}{\pi T}\right)^{1/2} + O\left(\frac{1}{N^{1/2}}\right).
\end{equation}
The result in equation~(\ref{rw1}) is well known (see, for example, Ref.~\cite{hughes}).
\begin{itemize}
\item Sublinear growth
of $S_N$ with time $N$ signifies that each site visited by such a walk is most probably
visited many times. This oversampling is precisely the reason why
searching a target in a one-dimensional system by a Brownian search is not very efficient,
since the walker wastes a great deal of time revisiting sites that do not contain the target.
That is why, in fact, recourse has been made to L\'evy-based searches, since they reduce
oversampling and lead to stronger growth of $S_N$ with $N$.
\item The result in equation~(\ref{rw2}) is the same as that in equation~(\ref{rw1}) with
the replacement $N \to N/T$ and hence does not represent a good search strategy either - in fact, it
is worse. When relocations over a distance $L$ take unit time as do
nearest-neighbor steps, the results in equations~(\ref{rw1}) and (\ref{rw2}) coincide, as they should.
\item Note as well that in one dimension,
equation~(\ref{survival}) with $S_N$ in equation~(\ref{rw1}) defines the asymptotically exact
behavior of the non-detection probability of a  target which \textit{diffuses}
in the presence of a concentration of \textit{diffusive} searchers \cite{bray}.
Thus the asymptotic behavior
of $P_N$ is \textit{independent} of the target diffusion coefficient
(see also Refs.~\cite{we} and \cite{sub1,sub2,sub3,sub4,sub5} for more details).
\item The result in equation~(\ref{survival}) can be generalized
to the case of imperfect recognition of the target,
that is, when target recognition upon encounter occurs with probability
$p < 1$~\cite{3}.  In one-dimensional systems the leading
asymptotic behavior of $P_N$ is \textit{independent} of $p$
provided that $p > 0$, and thus is also described by equation~(\ref{survival}).
\end{itemize}

We seek a large-$N$ expansion of $S_N$, in which (to arrive at the correct optimization) it
is essential to retain not only the
leading divergent contribution as $N\to \infty$ but also, if present,
a constant $N$-independent correction term.
Turning to the limit $z \to 1^{-}$ ($N \gg 1$) and inverting equation~(\ref{s})
we find, after some rather tedious but straightforward calculations, that $S_N$ obeys
\begin{eqnarray}
\label{s1}
S_N = f_1 N^{1/2}  + f_2 + O\left(\frac{1}{N^{1/2}}\right),
\end{eqnarray}
where the coefficient $f_1$ is given by
\begin{eqnarray}
\label{s2}
f_1 &=& \left(\frac{8}{\pi} \frac{\tau + L^2}{\tau + T}\right)^{1/2}, \quad \alpha>0,\\
\label{s20}
  &=& \left(\frac{8}{\pi T}\right)^{1/2}, \quad \alpha\equiv 0.
\end{eqnarray}
The parameter
\begin{equation}
\tau \equiv \frac{\alpha}{1 - \alpha}
\end{equation}
is an important physical parameter which defines a characteristic time for a ``continuous
tour of diffusion," that is, the typical time spent by a searcher on the substrate between two
consecutive off-lattice flights.  Note that the leading term in equation~(\ref{s1})
grows as $N^{1/2}$, which means that the leading
behavior is that of a random walk, albeit intermittent, unless there is an
additional dependence of $f_1$ on $N$ via the optimization of equation~(\ref{s1}) with respect
to $\alpha$ and $L$.  Note also that the
coefficients $f_1$, $f_2, \ldots$ are discontinuous functions of
$\alpha$, and that $\alpha \equiv 0$ is a singular point since it is not possible
for a random walker that skips over $L$ sites at each step to ever visit all
sites, cf. Ref.~\cite{lindenberg}.
This discontinuity should be viewed with appropriate caution since for any fixed finite $N$,
$S_N$ is a smooth function of $\alpha$; the discontinuity
arises because we are describing an asymptotic behavior that is strictly valid
only in the $N \to \infty$ limit. Additional details and
explanations can be found in Refs.~\cite{1} and~\cite{lindenberg}.

Returning to the asymptotic $N\to \infty$ limit,
for $\alpha \equiv 0$ the constant term $f_2 \equiv 0$, while for $0 < \alpha < 1$ (note
that this double-sided inequality is strict) it is determined by
\begin{eqnarray}
\label{comp}
f_2 &=& - \frac{2 \left(\alpha + (1 - \alpha) L^2\right)}{\pi \sqrt{\alpha (1 - \alpha)}} \int_0^{U_m} \frac{d u}{{\rm sh}(u) \sqrt{1 - \tau {\rm sh}^2(u)}} \nonumber\\
 && \times \left(\frac{1}{2} \frac{{\rm sh}^2(2 L u)}{{\rm sh}^2(L u) + \tau {\rm sh}^2(u)} - \frac{L}{\tau + L^2} {\rm cth}(u)\right),
\end{eqnarray}
where
\begin{equation}
U_m = \frac{1}{2} \ln\left( \frac{2 - \alpha}{\alpha} + \sqrt{\left(\frac{2 - \alpha}{\alpha}\right)^2 - 1}\right).
\end{equation}
The integral in equation~(\ref{comp}) can not be performed in closed form, but its important
contribution to the problem can be estimated. Anticipating that
effective search strategies take place when
$2 U_m L \gg 1$ (see below),
we find that in this limit the leading behavior of $f_2$ is given by
\begin{equation}
\label{s3}
f_2 \sim - \frac{2}{\pi }
\left(\tau^{1/2} + \frac{L^2}{\tau^{1/2}}\right) g(L), \;\;\; 0 < \alpha < 1,
\end{equation}
where $g(L)$ is a slowly-varying function of $L$,
\begin{equation}
\label{s4}
g(L) = \ln\left(\frac{L}{(1 + \tau) (1 - \alpha)^{1/2}}\right) + 0.126 +
O\left(\frac{1}{L}\right).
\end{equation}
\emph{Equations}~(\ref{s1})-(\ref{s4}) \emph{constitute our main general result} and provide
the basis for the design of an optimal strategy through the choice of $\alpha$ and $L$.
Below we discuss such a design and show
that, indeed, the optimal search strategies fulfill the assumption $2 U_m L \gg 1$.

\subsection{Optimization: Tuning $\alpha$ at fixed $L$}

To highlight the optimization procedure, we start with the
case studied analytically and numerically in~\cite{1}, shown again here for completeness,
namely, we tune $\alpha$ but hold the relocation length $L$ fixed.

Note that $S_N$ defined by equations~(\ref{s1})-(\ref{s4}) is a non-monotonic function of the
characteristic diffusion time $\tau$.
Differentiating $S_N$ with respect to $\tau$ (discarding a weak dependence of $g(L)$ on $\tau$),
we find that the maximum of $S_N$ with respect to $\tau$ is
given implicitly as the solution of the equation
\begin{equation}
\label{min1}
\frac{\partial f_1}{d  \tau} \left(\frac{8 N}{\pi}\right)^{1/2} = \frac{1}{\pi \tau^{1/2}}
\left(1 - \frac{L^2}{\tau}\right) g(L),
\end{equation}
where
\begin{equation}
\frac{\partial f_1}{d  \tau} = \frac{1}{2} \left(\frac{1}{\left[(\tau + T) (\tau + L^2)\right]^{1/2}}
- \frac{(\tau + L^2)^{1/2}}{(\tau + T)^{3/2}}\right).
\end{equation}
The left-hand-side of equation~(\ref{min1}) diverges when $N \to
\infty$, which indicates that in this case with fixed $L$ the
optimal time $\tau$ of continuous tours of diffusion should tend to
zero. We find that to leading order in $N$ the optimal
$\tau=\tau_{opt}$ and hence, the optimal value $\alpha_{opt}$ of the
intermittency parameter $\alpha$, are given by
\begin{equation}
\alpha_{opt} \sim \tau_{opt} \sim T \frac{L^{2/3} \ln^{2/3}(L)}{(2 \pi N)^{1/3}}.
\end{equation}
This is consistent with the condition $2 U_m L \gg 1$ since $U_m L
\sim L \ln(1/\alpha) \sim L \ln(N) \gg 1$. The symbol $\sim$ here and
henceforth signifies the exact behavior to leading order in $N$.
The expected number of distinct sites visited by an intermittent random
walk with an optimal $\alpha$ and fixed $L$ then is
\begin{equation}
S_N \sim \frac{L}{T^{1/2}} \left(\frac{8 N}{\pi}\right)^{1/2},
\end{equation}
i.e., it differs by a factor of $L/\sqrt{T}$ from the
corresponding result for a standard nearest-neighbor random walk with
$\alpha = 1$ (Brownian search), equation~(\ref{rw1}).

The essential result of this subsection is an enhancement by a factor
$L/\sqrt{T}$ of the expected number of distinct
sites visited by an intermittent random walk with the
distribution in equation~(\ref{Int}) and an appropriate $N$-dependent
choice of the intermittency parameter compared to the outcome of an
ordinary random walk.  Note that this effect appears in an
\textit{exponent} in the non-detection probability, so it can become
dramatically apparent. For example, for $L = 5$, $T = 1$ and $N =
10^4$, with a density of searchers as low as $\rho = 0.01$, the
non-detection probability for a Brownian search is $P_N \approx 0.2$
while that of the optimal intermittent search
($\alpha_{opt} \approx 0.07$ for these parameters) we find $P_N \approx
0.0003$, a reduction of three orders of magnitude!
Note finally that in one dimension for fixed $L$ the optimal
strategy involves a progressively smaller fraction of
nearest-neighbor steps as $N$ is increased.

\subsection{Optimization: Tuning $\alpha$ and $L$ for $T = 1$}

Next we consider both $\alpha$ and $L$ in equation~(\ref{Int}) to be tunable, but we fix $T=1$, that is,
relocation to a nearest neighbor and to a neighbor a distance $L$ away both take one unit of time.
This causes the relocation velocity $V=L/T$ to become dependent on $N$ via $L$.
Setting $T = 1$ in equations~(\ref{s1})-(\ref{s4}), we have
\begin{eqnarray}
\label{40}
\fl S_N = \left(\alpha + (1 - \alpha) L^2\right)^{1/2} \left(\frac{8 N}{\pi}\right)^{1/2} -
\frac{2}{\pi} \left(\left(\frac{\alpha}{1 - \alpha}\right)^{1/2} + \left(\frac{1 -
\alpha}{\alpha}\right)^{1/2} L^2 \right) g(L).
\end{eqnarray}
Note that the first term on the right hand side of equation~(\ref{40}) grows with $L$
and thus favors high values of $L$, but that
the second term is negative and contains a higher power of $L$.  This
``competition" suggests that there exists an optimal $N$-dependent value of $L$
which leads to a maximum in the number of distinct sites visited.

To determine the optimal value of $L$, we differentiate $S_N$ with
respect to $L$. Again discarding a logarithmically weak dependence
of $g(L)$ on $L$ and anticipating that the optimal value of $L$ is
large, such that $L \gg (\alpha/(1 - \alpha))^{1/2}$ (to be checked later for consistency),
we find that the optimal value $L=L_{opt}$ obeys
\begin{equation}
\label{length}
L_{opt} \ln(L_{opt}) \sim \left(\frac{\pi \alpha N}{2}\right)^{1/2},
\end{equation}
and hence,
\begin{equation}
\label{opt1}
L_{opt} \sim 2 \frac{(\pi \alpha N/2)^{1/2}}{\ln(\pi
\alpha N/2)}.
\end{equation}
Substituting this result into equation~(\ref{40}) then leads to the
expected number of distinct sites visited in an intermittent random
walk with an optimal length $L_{opt}$ of the relocation stage,
\begin{equation}
\label{mm}
S_N \sim 2 \left(\alpha (1 - \alpha)\right)^{1/2}
\frac{N}{\ln(\pi \alpha N/2)}.
\end{equation}
Note the parabolic form of the prefactor as a
function of $\alpha$. The prefactor vanishes in the pure
limits $\alpha = 0$ and $\alpha = 1$, indicating that optimization
in these pure cases is not possible. This is a reflection of the fact that the
leading behavior of $S_N$ in these limits is determined by terms proportional to $N^{1/2}$ and constant, independent of $N$
terms are absent.

Further optimizing the prefactor in equation~(\ref{mm}) with respect to $\alpha$,
we find that $\alpha_{opt} \to 1/2$ as $N \to \infty$, and hence
\begin{equation}
\label{res}
S_N \sim \frac{N}{\ln(\pi N/4)}.
\end{equation}
This result is
consistent with the earlier assumptions $2 U_m L \gg 1$ and $L\gg [\alpha(1-\alpha)]^{1/2}$ since here
$U_m$ and $\alpha/(1-\alpha)$ are constants while $L_{opt}$ diverges as $N \to \infty$.

The result~(\ref{res}) shows that when $T = 1$, optimization of the intermittent search
with respect to both $\alpha$ and
$L$ leads to an additional factor $N^{1/2}/\ln(N)$ in $S_N$ via the
coefficient $f_1$ in equation~(\ref{s1}), which results in a much
stronger dependence of the expected number of distinct sites visited
on the maximal time $N$. In fact, we have obtained
a behavior close to that of a two-dimensional Brownian motion, which
signifies that optimal intermittent random walks lead to only marginal
oversampling. The optimal strategy here consists of taking $\alpha_{opt}
= 1/2$ $(\tau_{opt} = T = 1)$, and relocation length $L_{opt}$ as given in
equation~(\ref{opt1}). Note that a similar result, i.e., that
the maximum $S_N$ is attained when the time spent on relocations is equal
to the time spent in the diffusive stage, has been obtained for a
model describing a stochastic search of a target site on a DNA by a
protein~\cite{berg,coppey}.

At this point one might be tempted to say that exactly the same
temporal behavior of $S_N$ as in equation~(\ref{res}) can be found without
resorting to any optimization procedure but by merely taking a L\'evy
walk with $\mu = 1$ in equation~(\ref{levy}) (Cauchy distribution).
Indeed, in this case one obtains (see equation~(2.20) in~\cite{gillis})
\begin{equation}
\label{cauchy}
S_N^{Cauchy} \sim \frac{3}{2\pi^2} \frac{N}{\ln(N)},
\end{equation}
where the normalization constant $B$ in equation~(\ref{levy}) with $\mu =
1$ has been set to $3/\pi^2$.  Remarkably, comparing the prefactors in
equations~(\ref{res}) and (\ref{cauchy})
one notes that the search based on the intermittent strategy with fine tuning of
the optimization parameters outperforms the one based on a
Cauchy distribution  due to a numerical factor which is more than six times larger
in the intermittent walk.

Moreover, we emphasize that these algorithms are very different in
their quality of exploration of space. Consider, for example, the
``density of visited sites" defined by
\begin{equation}
\Omega_N = \frac{S_N}{2M_N},
\end{equation}
where $M_N$ is the expected maximal displacement in, say, the positive direction
so that $2M_N$ is a measure of the range of the walk. The
parameter $\Omega_N$ is thus a measure of how many sites have been visited within
the range of the walk. By definition, $0 \leq \Omega_N \leq 1$.
For a L\'evy walk with a Cauchy distribution of the relocation length
the distribution of the maximal displacement is well-known~\cite{darling};
the tail of this distribution exactly follows the
behavior of the parent Cauchy variables and hence, $M_N^{Cauchy}$ is
infinite. This implies that the density of visited sites vanishes, $\Omega_N
= 0$, and thus the exploration quality is very poor.
The expected maximal displacement of the intermittent random walk
can be found from the general result of Ref.~\cite{comtet},
\begin{equation}
\label{range}
2M_N = \left(\alpha + (1 - \alpha) L^2\right)^{1/2} \left(\frac{8 N}{\pi}\right)^{1/2} + \gamma_N + O\left(\frac{1}{N^{1/2}}\right),
\end{equation}
where
\begin{equation}
\gamma_N = \frac{2}{\pi} \int^{\infty}_0 \frac{d k}{k^2} \ln\left(\frac{2 \left(1 - \alpha \cos(k) -
(1 - \alpha) \cos(k L)\right)}{\left(\alpha + (1 - \alpha) L^2 \right) k^2}\right).
\end{equation}
We thus find that
$\Omega_N \to 1/2$ as $N \to \infty$, which signifies that
intermittent random walks
have a very good exploration quality in that they visit
half of the sites within their range.
This is not a result expected \textit{a priori} since
we are dealing with random walks that involve steps not only to nearest neighbor sites but
also to distant sites.  We note that the exploration quality may be further enhanced
by optimizing both $S_N$ and $\Omega_N$.

\subsection{Optimization: Tuning $\alpha$ and $L$ for fixed $V$}

We finally turn to the most difficult case, when relocation over distance $L$ proceeds with a
finite fixed velocity $V$. Note that this differs from the previous case, where $T=L/V$ was fixed.
We take note of two points:
\begin{itemize}
\item We expect that $\alpha_{opt} \sim 1$ since flights over distance $L$ are less favorable
now that each relocation costs time $T=L/V$ during which no new sites are visited. If
it turns out that the optimal relocation distance again grows with $N$ (as it does, see below),
then the time $T$ grows with $N$ as well.  This in turn implies that
it might become more advantageous to remain on the lattice, which means that
the optimal $\tau$ might be larger than in the previous case.
\item On the other hand, the expression in equation~(\ref{s3}) is only valid when $2 U_m L \gg 1$. If
$\alpha_{opt} \sim 1$, then $U_m \sim \sqrt{(1 - \alpha)/\alpha} = 1/\sqrt{\tau} \ll 1$.
Consequently, approximation~(\ref{s3}) will be valid if the optimal characteristic time
$\tau$ is not too large, i.e.,  $\tau \ll L^2$.
\end{itemize}

Differentiating equation~(\ref{s1}) with respect to $\tau$ (again discarding the logarithmically
slow variation of $g(L)$ with $\tau$), we have
\begin{eqnarray}
\label{N5}
\fl \left(\frac{1}{\sqrt{(\tau + L/V)(\tau + L^2)}} - \frac{\sqrt{\tau + L^2}}{(\tau + L/V)^{3/2}}\right)
\left(\frac{8 N}{\pi}\right)^{1/2} = \frac{2}{\pi} \left(\frac{1}{\sqrt{\tau}} -
\frac{L^2}{\tau^{3/2}}\right) g(L).
\end{eqnarray}
Since equation~(\ref{s3}) is only
valid when $\tau \ll L^2$, equation~(\ref{N5}) can be simplified to yield
\begin{eqnarray}
\label{N6}
\frac{L}{(\tau_ + L/V)^{3/2}} \left(\frac{8 N}{\pi}\right)^{1/2} = \frac{2}{\pi}
\frac{L^2}{\tau^{3/2}} g(L).
\end{eqnarray}
Now we proceed as follows. We first assume that $\tau \ll L/V$,
find $\tau_{opt}$ and $L_{opt}$, and ascertain whether the assumption is valid.
Then we will follow with a more accurate calculation.

If $\tau \ll L/V$, then
\begin{equation}
\tau_{opt} = \frac{L^{5/3} g^{2/3}(L)}{V \left(2 \pi N\right)^{1/3}}.
\end{equation}
Substituting this expression into equation~(\ref{s1}), we have
\begin{equation}
S_N = V^{1/2} L^{1/2} \left(\frac{8 N}{\pi}\right)^{1/2} - \frac{2}{\pi} V^{1/2}
L^{7/6} g^{2/3}(L)
\left(2 \pi N\right)^{1/6}.
\end{equation}
Differentiating the latter expression with respect to $L$, we find that $L_{opt}$ is defined implicitly by
\begin{equation}
L_{opt}^{2/3} g^{2/3}(L_{opt}) = \frac{3}{7} \left(2 \pi N\right)^{1/3},
\end{equation}
which implies that
\begin{equation}
\tau_{opt} = \frac{L_{opt}}{V} \frac{L_{opt} g^{2/3}(L_{opt})}{\left(2 \pi N\right)^{1/3}}
 = \frac{3}{7} \frac{L_{opt}}{V}.
\end{equation}
Hence, $\tau_{opt}$ is \emph{not} much smaller than $L_{opt}/V$.  It is only
smaller by a numerical factor and scales
as $L_{opt} \sim N^{1/2}/\ln(N)$ (note that nonetheless $\tau_{opt} \ll L_{opt}^2$). Note also that
$\tau_{opt} \to \infty$ as $N \to \infty$, which implies that
$\alpha_{opt} \to 1$ and $U_m \sim (1 - \alpha)^{-1/2} \sim \tau^{1/2}$.

We next try to search for optimal $L$ and $\tau$ from equation~(\ref{N6}) supposing that
$U_m \sim \tau^{1/2}$ and $\tau = C L/V$, where $C$ is a constant to be determined.
We note that in this case
\begin{eqnarray}
g(L) &=& \ln(L) - \ln\left((1 + \tau) \sqrt{1 - \alpha}\right) +
0.126 + O\left(\frac{1}{L}\right)\nonumber\\
&\approx& \frac{1}{2} \ln(L) - \frac{1}{2} \ln(C/V) \nonumber\\
&\approx& \frac{1}{2} \ln(L).
\end{eqnarray}
Hence, equation~(\ref{N6}) becomes
\begin{eqnarray}
\label{N7}
\frac{V^{3/2} L}{(1 + C)^{3/2} L^{3/2}} \left(\frac{8 N}{\pi}\right)^{1/2} = \frac{2}{\pi}
\frac{L^2}{\tau^{3/2}} \frac{1}{2} \ln\left(L\right).
\end{eqnarray}
From equation~(\ref{N7}) we find that the optimal value of $\tau$ obeys
\begin{equation}
\tau_{opt} =
 \frac{(1 + C)}{V} \frac{ L^{5/3} \ln^{2/3}\left(L\right)}{\left(2 \pi N\right)^{1/3}},
\end{equation}
and $S_N$, optimized with respect to $\tau$, then follows:
\begin{equation}
S_N = \left(\frac{L V}{1 + C}\right)^{1/2} \left(\frac{8
N}{\pi}\right)^{1/2} - \frac{1}{\pi}
\left(\frac{V}{1 + C}\right)^{1/2} L^{7/6}
\ln^{2/3}\left(L\right) \left(8 \pi N\right)^{1/6}.
\end{equation}
Differentiating this equation with respect to $L$, we find that the optimal flight length $L$ obeys
\begin{equation}
L_{opt} \ln\left(L_{opt}\right) =
\left(\frac{3}{7}\right)^{3/2} \left(8 \pi N\right)^{1/2},
\end{equation}
so that
\begin{equation}
\label{lopt}
L_{opt} \sim  \frac{\left(\frac{3}{7}\right)^{3/2} \left(8
\pi N\right)^{1/2}}{\ln\left(
\left(\frac{3}{7}\right)^{3/2} \left(8 \pi
N\right)^{1/2}\right)}.
\end{equation}

Lastly, we obtain $C$ from the definition $\tau_{opt} = C L_{opt}/V$. This gives
\begin{equation}
(1 + C) \frac{L_{opt}}{V}
\frac{L_{opt}^{2/3} \ln^{2/3}\left(L\right)}{\left(8
\pi N\right)^{1/3}} = C \frac{L_{opt}}{V}
\end{equation}
which is solved by
\begin{equation}
C = \frac{3}{4}.
\end{equation}
Thus the optimal strategy is realized when
we choose $L_{opt}$ as in equation~(\ref{lopt}), and $\tau = (3/4) T$, i.e., for the optimal strategy
the characteristic time of a tour of diffusion between two consecutive long jumps is three-fourths
of the time a searcher spends on jumps over a distance $L$.

Finally, we combine these results to find that the expected number of distinct sites
visited optimized with respect to both $\alpha$ and $L$ at fixed $V$ is
\begin{equation}
S_N = \left(\frac{4}{7}\right)^{1/2} V^{1/2} L_{opt}^{1/2} \left(\frac{8 N}{\pi}\right)^{1/2} \sim
V^{1/2} \frac{N^{3/4}}{\ln^{1/2}(N)}
\end{equation}
Note that this more intricate optimization procedure results in a
faster growth law of $S_N$ with $N$ than in the pure random walk cases.
Consequently, the search efficiency has again been enhanced by orders
of magnitude, albeit not as much as in the fixed-$T$ case of the previous subsection.

\section{Conclusions}

We have considered the design of an optimal search strategy of a hidden target
by a given density $\rho$ of random walkers who have a limited maximal time $N$ to find
the target.  The analysis is restricted to one-dimensional systems.
Our measure for the quality of a strategy is
\emph{the minimization of the probability} $P_N$ \emph{that the target is undetected within
the given maximal search time}. In particular, we consider strategies that consist of
a combination of nearest neighbor walks and jumps of fixed length $L$, both involving steps in
either direction with equal probability.  The motion of each searcher is thus intermittent.
The probability that the target is undetected is just the survival probability for the target and is
related to the distinct number of sites $S_N$ visited by a walker up to time $N$ by the
well-known relation $P_N = \exp(-\rho S_N)$.  Our goal has thus been to maximize $S_N$. We
stress that the time derivative of $P_N$ defines the distribution function of the
first passage time to the detection event. This means that in contrast to previous
work ~\cite{benichou1,benichou2,benichou3,coppey,1},
our aim has been to optimize the full distribution function and not
only its first moment.

Our model has three parameters: $\alpha$, the probability that the next step of the walker is a
nearest neighbor step; $L$, the length of a long step; and $T=L/V$, where $T$ is the time it
takes to cover a long step and $V$ is the velocity of a long step. The parameters $\alpha$ and $L$
are optimized as a function of the maximal time $N$ under different constraints.
We compare our results for $S_N$ with those of a nearest neighbor random walk, $S_N \sim
(8N/\pi)^{1/2}$, and of a L\'evy walk with a Cauchy distribution of step lengths, $S_N\sim
(3/2\pi^2) N/\ln(N)$.

If $L$ and $T$ are fixed and only $\alpha$ is picked for optimal strategy, the best choice is to
take it to be very small, $\alpha \sim N^{-1/3}$~\cite{1}. Most of the random motion then consists of
steps of length $L$. The resulting $S_N$ is larger than that
obtained with a nearest neighbor random walk by the numerical coefficient $L/\sqrt{T}$,
$S_n\sim (L/T^{1/2}) (8N/\pi)^{1/2}$, so that the
non-detection probability, which involves this factor in the exponent, can be decreased
by orders of magnitude even for a low density of searchers.

If we optimize both $\alpha$ and $L$,
the best choice of these parameters depends on the constraint we place on the third parameter.
It the time for a long step is the same as the time for a nearest neighbor step, then the optimal
choices are $\alpha = 1/2$ and $L \sim N^{1/2}/ln(N)$. Thus, short and long steps should occur with
equal likelihood, and the optimal distance covered by the long steps grows (slowly) with increasing
observation time.  In this case the distinct number of sites
visited is even larger (again by a numerical factor) than that of a L\'evy walk, $S_N \sim
N/\ln(\pi N/4)$. These $N$ dependences are the same for any fixed value of $T$, although the
specific numerical coefficients depend on this value.
We find an important difference between the L\'evy walk and our optimized
intermittent walk in the coverage of space, which may be an important feature if one wishes to
parallelize the searches of different walkers. The density of visited sites vanishes in the L\'evy
case, while that of the intermittent walk approaches a constant with increasing $N$.

Finally, if we again optimize both $\alpha$ and $L$ but now keeping the velocity $V$ of the long
steps fixed (which means that the time for a long step grows with the length of the step), we find
the optimal choice $\alpha$ to be close to unity, $\alpha \sim
1 - 4 V ln(N)/3 N^{1/2}$, and the optimal length step to grow with $N$ as
$L \sim N^{1/2}/ln(N)$.  Now the walkers rarely jump over long distances, but the time spent
diffusing and the time it takes to make a jump of length $L$ both grow with increasing observation
time $N$, so that $\tau = 3T/4$. The distinct number of sites visited here shows a growth
intermediate between the other two intermittent strategies,
$ S_N \sim V^{1/2} N^{3/4}/\ln^{1/2}(N)$.

We stress that in all the strategies considered here, we have implemented a maximum observation time
$N$ as part of the optimization process,  and we have used a particular measure of the quality of a
strategy, namely, that the probability of survival of the target be minimized within this time.
We have shown that in one dimension even a simple intermittent step
distribution consisting of nearest neighbor steps and long steps of fixed $N$-dependent
length can be optimized so as to yield far better search outcomes than a nearest neighbor random
walk and even a L\'evy walk with a Cauchy distribution of step lengths.  The model can be
generalized and further optimized in a number of ways, some of which we have noted in the course of
our analysis.

\subsection{Acknowledgments}

The authors gratefully acknowledge helpful discussions with O.B\'enichou, J. Klafter and A. Grosberg.
The research of GO is partially supported by Agence Nationale de la Recherche (ANR) under grant
``DYOPRI - Dynamique et Optimisation des Processus de Transport Intermittents".
HSW acknowledges partial support from MEC, Spain, through Grant CGL2007-64387/CLI, and from AECID, Spain, project A/018685/08.

%% Add your bibliography items here.  PNAS requires that bibliography items
%% be entered directly into the article rather than called from a BibTeX
%% environment.  Contact pnas@nas.edu if you need assistance with your
%% bibliography.

% Sample bibliography item in PNAS format:
%% \bibitem{Ch} D. Chae (2003) {\it Nonlinearity} {\bf 16}, 479-495.
%% \bibitem{in-text reference} Author Names (year published)
%% {\it Journal Name} {\bf Volume #}, start page-end page

\end{document}